\documentclass[%
reprint,
superscriptaddress,
 amsmath,amssymb,
 aps,
pra,
]{revtex4-1}

\usepackage{graphicx}
\usepackage{dcolumn}
\usepackage{bm}
\usepackage{physics}
\usepackage{siunitx}
\usepackage{mathtools}
\usepackage{extarrows} 

\usepackage{nth}

\usepackage{dsfont}

\usepackage{xr-hyper}
\usepackage{hyperref}
\hypersetup{
    colorlinks = true,
    allcolors = {blue},
    urlcolor = {black}
}

\usepackage{xfrac}

\usepackage{xcolor} 

\usepackage{soul}
\usepackage{bigints}
\usepackage{subfigure}
\usepackage{chemformula}
\usepackage{lipsum} 
\usepackage[capitalize]{cleveref}

\NewDocumentCommand{\evalat}{sO{\big}mm}{%
  \IfBooleanTF{#1}
   {\mleft. #3 \mright|_{#4}}
   {#3#2|_{#4}}%
}

\newcommand{\appropto}{\mathrel{\vcenter{
  \offinterlineskip\halign{\hfil$##$\cr
    \propto\cr\noalign{\kern2pt}\sim\cr\noalign{\kern-2pt}}}}}

\begin{document}

\title{Scheme for single-shot frequency comb absorption sensing on chip}

\author{Jake Biele}
\email{jake.biele7@gmail.com}
\affiliation{%
Quantum Engineering Technology Labs, H. H. Wills Physics Laboratory and Department of Electrical \& Electronic Engineering, University of Bristol, Bristol BS8 1FD, United Kingdom
}%
\affiliation{%
Quantum Engineering Centre for Doctoral Training, University of Bristol, Bristol BS8 1FD, United Kingdom
}%
\author{Sabine Wollmann}
\email{s.wollmann@hw.ac.uk}
\affiliation{%
Quantum Engineering Technology Labs, H. H. Wills Physics Laboratory and Department of Electrical \& Electronic Engineering, University of Bristol, Bristol BS8 1FD, United Kingdom
}%
\affiliation{%
School of Engineering and Physical Sciences, Heriot-Watt University, Edinburgh, EH14 4AS, UK, 
}
\author{Euan Allen}
\email{ea901@bath.ac.uk}
\affiliation{%
Quantum Engineering Technology Labs, H. H. Wills Physics Laboratory and Department of Electrical \& Electronic Engineering, University of Bristol, Bristol BS8 1FD, United Kingdom
}%
\affiliation{%
Centre for Photonics and Photonic Materials, Department of Physics, University of Bath, Bath, BA2 7AY, UK}

\date{\today}

\begin{abstract}
Frequency comb absorption spectroscopy combined with low-noise, fast homodyne measurements provide a toolbox for ultra-sensitive absorption measurements. Integrating these schemes on photonic platforms to bring them closer to practical applications is challenging. Here, we propose a scheme that can be readily adapted on a photonic platform. 
We show that each frequency comb tooth can independently sampled, the span of the comb can be tailored, and that our method does not require a careful alignment of the comb centre with any absorption profiles. This allows an asymmetric absorption profile to be reconstructed in full without requiring additional components. 
\end{abstract}
\maketitle
Spectroscopy allows us to study and monitor material to enhance our understanding of its composition and structure. This field has tremendously benefited from technological development and new sensing schemes over the past years. This has enabled practical applications of spectroscopy in areas ranging from environmental monitoring~\cite{vlk2021extraordinary, ycas2019mid} to industrial applications~\cite{Lackner2007Apr,linnerud1998gas}. These applications mostly use a tunable, narrow-linewidth laser in combination with direct absorption schemes which provides a reconfigurable toolbox to address individual spectral absorption lines~\cite{duarte2008tunable, du2019mid}. However, the sensitivity and precision is ultimately limited by noise and power fluctuations of the laser~\cite{Hodgkinson2012Nov}. To avoid frequency scanning of the light source to match individual absorption lines, a multi-spectral frequency comb can be used.\\ 
Frequency combs (FC) are generated by modulating a continuous-wave laser and have equidistant, coherent spectral sidebands, so-called comb teeth~\cite{Bjorklund1980Jan}. This gives access to a probe with a much wider spectral range and the narrow line-width of the teeth significantly improves the spectral resolution of the absorption measurement. Schemes such as heterodyne (HS) and dual-comb spectroscopy (DCAS) provide a fast and sensitive toolbox for multi-wavelength measurements~\cite{abbas2019time, Picque2019Mar}. These spectral properties, the broadband spectrum with simultaneous coherence between the comb teeth, benefit trace gas detection schemes that can improve their accuracy of concentration measurements and  show a stronger noise robustness to coupling and phase drifts thanks to the increase of measurement points per teeth.  
However, these sensing schemes still demand sophisticated stabilisation and locking schemes and rely on bulky and lossy high-frequency modulation components. To bring them closer to practical applications, they have to be integrated on stable, low-cost, and scalable platforms. This makes silicon photonics a strong candidate to address these requirements. Further, the cheap large-scale manufacturing combined with high reconfigurability~\cite{qiang2018} makes them suitable for realising practical applications, such as lab-on-chips or environmental monitors.
However, integrating existing schemes on such platforms is challenging. DCAS requires coherence between the two interfering FCs that can be achieved by tuning the spectra of one of the FC with respect to the other. This requires independent temperature tuned photon sources on chip which is non-trivial to achieve. Further, low optical signals, e.g. $\sim 50\%$ in HS detection, can reduce the signal-to-noise ratio of measurements, require additional lasers or lossy acousto-optic modulators to be integrated.

\begin{figure}[htb]
    \centering
    \includegraphics[width=\linewidth]{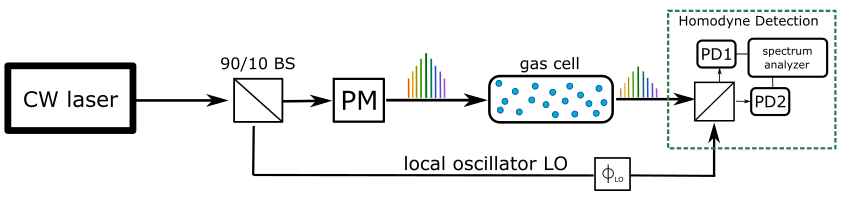}
    \caption{Scheme for frequency comb sensing. A frequency comb is generated by phase modulating (PM) a coherent state generated by a CW laser. This probe is measuring a gas with frequency-dependent transmission $\eta$. The detection system consists of balanced homodyne detection consisting of two photodiodes, PD1 and PD2, a local oscillator (LO) with tunable phase $\phi_{LO}$ followed by a spectrum analyzer.}
    \label{fig:sketch}
\end{figure}

Here, we propose a novel sensing scheme that can be readily implemented on photonic platforms. Further, this strategy combines the advantages of a strongly modulated, broad FC with fast and sensitive homodyne measurements (Fig.~\ref{fig:sketch}). This requires only one optical input for the laser, whilst all other integrated components, such as detectors and modulators, can be electrically controlled up to the GHz regime~\cite{Tasker2021Jan}.
Besides the technological impact, we will show how our scheme allows us to maintain the advantages of homodyne detection without having to make assumptions about the distribution of loss or absorption across the positive and negative optical sideband which typically demands a well-centered FC with respect to the absorption profile of a gas.
Our scheme allows us to overcome these limitations by controlling the local oscillator phase when performing homodyne tomography.  By measuring two orthogonal quadratures,a one-to-one mapping of losses on the optical sidebands can be extracted whilst maintaining the superior SNR of homodyne detection. Additionally, we further reduce assumption about the modulation depth that controls the span width of the FC and present a detailed analysis how to extract the absorption loss on each sideband.

--\textit{Weak modulation.}\\
The sensing strategy we consider is based on a coherent-state $\ket{\alpha}$ that is modulated to produce a bright frequency comb (FC) for our absorption measurement (Fig.~\ref{fig:sketch}). Our FC interacts with a sample undergoing a frequency-dependent absorption before it interferes with a bright local oscillator (LO). The measurement consists of homodyne detection and a read-out, e.g. a spectrum analyzer.
The initial coherent state $\ket{\alpha}$ with a center frequency $\omega_0$ is modulated to generate optical sidebands at $\pm m\Omega$ from the carrier frequency $\omega_0$. We can either phase (pm) or amplitude modulate (am) with a sinusoidal frequency $\Omega$. Previous sensing schemes used weak modulation which results in the generation of $m=1$ sideband,
\begin{equation}
\begin{split}
     A_{pm}&=|\alpha| e^{i(\delta\sin(\Omega t)}\\
     A_{am}&=|\alpha| (1+\delta\sin(\Omega t)\\
\end{split}
\label{eq:modulation} 
\end{equation}
where $A_{pm}$ is our phase modulated signal, $A_{am}$ is our amplitude modulated signal.

If a direct detection with one photodiode is used, contributions of higher order terms in the monitored current $i$ become negligible.
\begin{equation}
\begin{split}
     i_{pm}(t)&=|\alpha| e^{-i(\delta\sin(\Omega t)}|\alpha| e^{i(\delta\sin(\Omega t)}=|\alpha|^2\\
     i_{am}(t)&=|\alpha|^2 (1+\delta\sin(\Omega t)^2=|\alpha|^2+2|\alpha|^2\delta \sin(\Omega t)
\end{split}
\end{equation}
We use a spectrum analyser with resolution bandwidth $B$ and input impedance R to measure the signal in the sidebands, which gives different results for phase and amplitude modulated FCs,

\begin{equation}
\begin{split}
     p_{pm}&=0\\
     p_{am}&=2R|\alpha|^4 \delta^2
\end{split}
\end{equation}
We can see that phase modulation does not contribute to the coherent signal in the sideband whereas amplitude modulation does lead to a signal - this is not surprising, weak phase modulation should not be detectable via an intensity measurement.
We now consider a homodyne detection scheme where we interfere our FC with a classical local oscillator (LO), described by $B=|\beta| e^{i\phi_{LO}}$ and its phase $\phi_{LO}$.

This leads to a detectable signal by the spectrum analyzer of
\begin{equation}
\begin{split}
    p_{pm}&=2R|\alpha|^2|\beta|^2\delta^2\sin^2(\phi_{LO})\\
    p_{am}&=2R|\alpha|^2|\beta|^2 \delta^2\cos^2(\phi_{LO})
\end{split}
\end{equation}
When $\phi_{LO}=0$, i.e. we measure the amplitude quadrature, the phase modulation again does not contribute to the signal. Shifting the LO to measure the phase quadrature gives a signal from PM and not AM as expected.It was demonstrated by one of us and colleagues~\ref{George}  that such modulation leads to improved signal-to-noise ratio in absorption measurements.\\
--\textit{Strong modulation.}\\
Extending the previous work to the strong modulation regime, allows for generation of FC with multiple teeth and significant wider spectral span. This allows to resolve multiple absorption features of a gas whilst simultaneously improving the measurement accuracy as each tooth provides a frequency-dependent absorption signal. Further, we will see that our scheme allows us to distinguish between the absorption of the positive and negative sideband whilst removing the requirement to center the FC with the absorption profile.\\
A strong signal modulation of depth $\delta$ allows for a generation of $m$ sidebands in the signal which limits the span of the FC to $\omega_0\pm m\Omega$. The phase modulated FC can be generally described using the Jacobi-Anger expansion for the signal $\hat{A}$.
\begin{equation}
\begin{split}
   \hat{A} &= |\alpha|\Bigg(\mathrm{J}_0(\delta)+\sum_{m=1}^{\infty}\mathrm{J}_m(\delta)e^{im\Omega t}+\sum_{m=1}^{\infty}(-1)^m\mathrm{J}_m(\delta)e^{-im\Omega t}\Bigg),
\end{split}
\end{equation}
Here, $J_m(\delta)$ are the Bessel functions of order $m$ with modulation depth $\delta$~\cite{Bjorklund1980Jan}.\\
We consider this phase-modulated signal $A$ to be our probe field for sensing a gas. We assume our gas to be dispersion-less and to be weakly absorbing, hence to give a negligible phaseshift to our probing signal. However, frequency-dependent phaseshifts that apply small offsets to the quadratures of each comb tooth when considering practical scenarios. We will discuss the impact of this further in a later section.
Applying the Beer-Lambert model of loss to each comb tooth, we can define the frequency-dependent transmission of each optical sideband through the sample of length L as $\eta_{+m}=e^{V(\omega_0+m\Omega)L}$.  The probe signal after its interaction with a gas can be described as,
\begin{equation}
\begin{split}
 \hat{A} &= |\alpha|\Bigg(\sqrt{\eta_0}\mathrm{J}_0(\delta)+\sum_{m=1}^{\infty}\sqrt{\eta_{+m}}\mathrm{J}_m(\delta)e^{im\Omega t}\\
 &+\sum_{m=1}^{\infty}(-1)^m\sqrt{\eta_{-m}}\mathrm{J}_m(\delta)e^{-im\Omega t}\Bigg)
\end{split}
\end{equation}
This leads to a photocurrent in our homodyne detection of the form:
\begin{equation}
\begin{split}
\hat{i}(t) &= 2|\alpha||\beta|\Bigg[\sqrt{\eta_0}\mathrm{J}_0(\delta)\cos(\phi_{\mathrm{LO}})\\
     &+ \sum_{m=1}^{\infty}\mathrm{J}_m(\delta)\cos(\phi_\mathrm{LO})\cos(m\Omega t) \Big[\sqrt{\eta_{+m}}+(-1)^m\sqrt{\eta_{-m}}\Big] \\
    &+ \sum_{m=1}^{\infty}\mathrm{J}_m(\delta)\sin(\phi_\mathrm{LO})\sin(m\Omega t) \Big[\sqrt{\eta_{+m}}-(-1)^m\sqrt{\eta_{-m}}\Big] \Bigg]
\end{split}
\end{equation}

We take the Fourier transform, integrate about the $m^{\mathrm{th}}$ tooth $\int^{m\Omega}$, and take the modulus squared to give a final expression for the power about the $m^\mathrm{th}$ tooth as measured by a spectrum analyser:

\begin{equation}
    \begin{split}
        p_{m\Omega}&=2R|\alpha|^2|\beta|^2\mathrm{J}_m^2(\delta)\Bigg[\cos^2(\phi_\mathrm{LO})
        \Big[\sqrt{\eta_{+m}}+(-1)^m\sqrt{\eta_{-m}}\Big]^2 \\
        &+ \sin^2(\phi_\mathrm{LO})\Big[\sqrt{\eta_{+m}}-(-1)^m\sqrt{\eta_{-m}}\Big]^2 \Bigg]
    \end{split}
\end{equation}

By sweeping the LO phase $\phi_{LO}$ and solving for the minimum and maximum of the resulting sine curve for each tooth, it is possible to extract $\eta_+$ and $\eta_-$. This means each frequency-dependent tooth can be independently measured using $\phi_{LO}$. In the special case that the transition is symmetrical, the signal is entirely in alternate quadratures dependent on the tooth order $m$. By introducing the LO phase and taking measurements across $0\leq \phi_{LO} \leq 2\pi$, we can map out the maximum and minimum signal for each tooth which in turn gives us a one-to-one mapping to calculate the target transition. In direct detection schemes, amplitude modulation is required and one would not be able to break the symmetry to give a one-to-one mapping. Conversely, in heterodyne detection, the symmetry is broken by offsetting the beat signals from each pair of sidebands but this comes at the additional cost of a drop in SNR of 3dB and requires either two lasers or another modulator on chip. In our measurement scheme, will scan the full $2\pi$ of LO phase and extract the transition. The noise floor of the measurement will determine the minimum difference in loss we can infer between each pair of teeth.\\
---\textit{Conclusion.}\\
Our novel sensing scheme simultaneously combines the advantages of FC sensing with fast homodyne measurements. This provides a sensing model that gives access to a phase-stable local oscillator that allows us to amplify the probing signal. Establishing such coherence is not trivial and requires additional stabilisation schemes in other schemes, e.g. dual-comb spectroscopy~\cite{Coddington2016Apr, Picque2019Mar}.
Moreover, measuring absorption features of gas sample at each frequency with independently measured comb teeth makes scanning of the laser redundant and provides a significant speedup in measurement time and additionally, makes the scheme more robust to environmental fluctuations in comparison to alternative spectroscopy schemes~\cite{Thorpe2006Mar,Diddams2007Feb,Karim2020Aug,Stowe2008Jan}.
Further, controlling the modulation depth allows us to customise the width of the FC to optimally resolve an absorption profile.
Overall, this theoretical model is readily adaptable for a photonic platform. Using state-of-the-art integrated frequency comb sources~\cite{Xiang2021Jul, Yang2021Aug}, low-loss waveguide-based sensing structures~\cite{Ji2017Jun, Belsley2022Jun}, followed by on-chip homodyne detection~\cite{Tasker2021Jan} has a strong potential to bring a compact and robust spectroscopic gas sensor closer to practical application.

\bibliography{mainbib.bib}

\end{document}